# Giant polarization charge density at lattice-matched GaN/ScN interfaces

Nicholas L. Adamski, Cyrus E. Dreyer, and Chris G. Van de Walle

View Online     Export Citation     CrossMark

**ARTICLES YOU MAY BE INTERESTED IN**

BAlGaN alloys nearly lattice-matched to AlN for efficient UV LEDs
Applied Physics Letters **115**, 231103 (2019); https://doi.org/10.1063/1.5129387

Large bandgap tunability of GaN/ZnO pseudobinary alloys through combined engineering of anions and cations
Applied Physics Letters **115**, 231901 (2019); https://doi.org/10.1063/1.5126930

Transmorphic epitaxial growth of AlN nucleation layers on SiC substrates for high-breakdown thin GaN transistors
Applied Physics Letters **115**, 221601 (2019); https://doi.org/10.1063/1.5123374

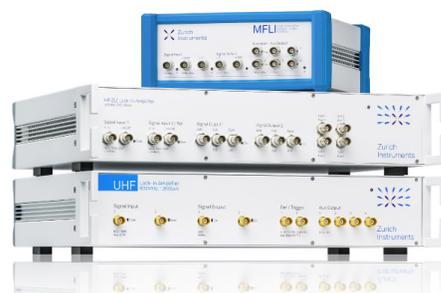

# Giant polarization charge density at lattice-matched GaN/ScN interfaces

Nicholas L. Adamski,[1,a]] 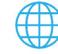 Cyrus E. Dreyer,[2] and Chris G. Van de Walle[3]

AFFILIATIONS

[1]Department of Electrical and Computer Engineering, University of California, Santa Barbara, California 93106-9560, USA
[2]Department of Physics and Astronomy, Stony Brook University, Stony Brook, New York 11794-3800, USA and Center for Computational Quantum Physics, Flatiron Institute (a division of the Simons Foundation), 162 5th Avenue, New York, New York 10010, USA
[3]Materials Department, University of California, Santa Barbara, California 93106-5050, USA

[a)]Electronic mail: nadamski@ece.ucsb.edu

ABSTRACT

Rock-salt ScN is a semiconductor with a small lattice mismatch to wurtzite GaN. Within the modern theory of polarization, ScN has a nonvanishing formal polarization along the [111] direction. As a result, we demonstrate that an interface between (0001) GaN and (111) ScN exhibits a large polarization discontinuity of $-1.358$ Cm$^{-2}$. Interfaces between ScN and wurtzite III-nitrides will exhibit a high-density electron gas at the $(000\bar{1})$ GaN interface or a hole gas at the (0001) GaN interface, with carrier concentrations up to $8.5 \times 10^{14}$ cm$^{-2}$. The large polarization difference and small strain make ScN a desirable choice for polarization-enhanced tunnel junctions within the III-nitride materials system. The large sheet carrier densities may also be useful for contacts or current spreading layers.

Published under license by AIP Publishing. https://doi.org/10.1063/1.5126717

ScN is a semiconducting nitride that takes the rock-salt (rs) crystal structure and can be integrated with the technologically interesting III-nitride family of compounds. Recently, a great deal of attention has been focused on alloys of ScN and AlN, which exhibit enhanced piezoelectricity at Sc concentrations up to 43%[1] and ferroelectricity at Sc concentrations between 27% and 43%.[2] Pure rock-salt ScN has attracted interest for its low lattice mismatch to GaN; grown along the [111] direction, it exhibits a mismatch of less than 1% with c-plane GaN.[3] ScN grown by molecular beam epitaxy has been examined as a potential buffer layer for improving the quality of heteroepitaxial GaN.[4,5] However, questions still remain about properties of the pristine interface, as electrical characterization has only been conducted on polycrystalline films[6] or on films with oxide contamination at the interface.[3]

Interfaces between wurtzite (wz) III-nitrides have charges induced by polarization differences.[7] These polarization charges are desirable in power electronics, where they result in large two-dimensional electron and hole gases (2DEGs and 2DHGs).[8,9] Polarization can also be utilized to enhance the field in tunnel junctions.[10–12] Current tunnel structures are based on a GaN p-n junction with a thin AlGaN or InGaN interlayer that creates a strong polarization field across the depletion region. The high Al or In content is desirable to increase the field strength and reduce the width of the depletion region, but the alloy composition is constrained by strain considerations.[13]

In this work, we use density functional theory and the modern theory of polarization to demonstrate that a large polarization sheet charge with a magnitude of 1.358 Cm$^{-2}$ will exist at the ScN/GaN interface. We clarify that the reason for this charge is not a result of the naïve expectation that, since rock salt (rs) is centrosymmetric (space group $Fm\bar{3}m$), it has no polarization,[14] while wurtzite (wz) GaN has a large spontaneous polarization. In fact, rs-ScN has a nonvanishing "formal" polarization[15] along the [111] direction, which must be taken into account when calculating the polarization differences that lead to bound charges at the interface. This is confirmed by comparing with explicit superlattice calculations. Based on the large polarization discontinuity, we will propose potential applications of ScN interfaces.

The density functional calculations are performed using within the Vienna *Ab initio* Simulation Package (VASP).[16,17] The projector augmented wave potentials[18] include N $2s^22p^3$, Sc $3d^14s^2$, and Ga $4s^24p^1$ electrons as valence electrons. We use the hybrid functional of Heyd, Scuseria, and Ernzerhof (HSE) with a standard mixing parameter of 25%.[19,20] The plane wave energy cutoff is 500 eV. Γ-centered $k$-point grids of size $15 \times 15 \times 15$ are used for bulk ScN calculations and $8 \times 8 \times 6$ for bulk GaN. Interfaces are modeled using superlattices containing 8 formula units each of GaN and ScN, with a $7 \times 7 \times 1$ $k$-point grid.

ScN has a rock-salt structure, which consists of two interpenetrating face-centered cubic lattices. Using the 2-atom primitive unit cell of ScN, we calculate the lattice parameter to be $a_{ScN} = 4.48$ Å (which corresponds to an in-plane wurtzite equivalent of 3.17 Å) with an indirect bandgap of 0.80 eV, in agreement with experimental values.[21] The lattice parameters of GaN are $a_{GaN} = 3.20$ Å and $c_{GaN} = 5.20$ Å, with a direct bandgap of 3.18 eV. The underestimation of the GaN bandgap (experiment: 3.51 eV, Ref. [22]) results from the necessity to choose a single HSE mixing parameter for the ScN/GaN superlattice calculations. This does not affect the quantities extracted from the superlattice calculations,[23] nor the accuracy of the calculated polarization values.

Within the modern theory of polarization, the property of focus is the formal polarization,[24,25]

$$\mathbf{P}_f = \frac{e}{\Omega} \sum_s Z_s \mathbf{R}_s + \frac{ief}{8\pi^3} \sum_j^{occ} \int_{BZ} d\mathbf{k} \langle u_{j,\mathbf{k}} | \nabla_{\mathbf{k}} | u_{j,\mathbf{k}} \rangle. \quad (1)$$

The first term is the ionic contribution, where $e$ is the electron charge, $\Omega$ is the volume of the unit cell, $Z_s$ is the charge of ion $s$, and $\mathbf{R}_s$ is the position of ion $s$ within the structure. The second term is the electronic contribution, where $f$ is the spin degeneracy of the bands, and $\langle u_{j,\mathbf{k}} | \nabla_{\mathbf{k}} | u_{j,\mathbf{k}} \rangle$ is the Berry potential, which is integrated over the Brillouin zone to give the Berry phase, summed over all the occupied bands $j$. Physically observable polarization properties are always determined from differences in formal polarization.[24,25]

The formal polarization of a material is defined only modulo the "quantum of polarization" $e\mathbf{R}/\Omega$, where $\mathbf{R}$ is any lattice vector. This multivalued vector (which leads to different "branches" in $\mathbf{P}_f$) must map onto itself under the symmetry operations of the crystal.[15] For rock-salt symmetry, there are two distinct sets of vectors that satisfy all the symmetry operations: the formal polarization can be written as $\frac{e}{\Omega}\mathbf{R}$ or $\frac{e}{\Omega}[\mathbf{R} + a_{rs}(\frac{1}{2}, \frac{1}{2}, \frac{1}{2})]$, where $a_{rs}$ is the rock-salt lattice parameter. From explicit evaluation of Eq. (1), we find that ScN has the formal polarization $\frac{e}{\Omega}[\mathbf{R} + a_{rs}(\frac{1}{2}, \frac{1}{2}, \frac{1}{2})]$.

Our goal is to determine the bound polarization charge at an interface between wz-GaN in the [0001] direction and rs-ScN in the [111] direction. For this purpose, we use the interface theorem:[26] if an insulating interface can be constructed between the two structures, the bound charge at the interface can be determined from the differences in the formal polarization,

$$\sigma_b = \left(\mathbf{P}_f^{GaN} - \mathbf{P}_f^{ScN}\right) \cdot \hat{n}. \quad (2)$$

The formal polarizations are only defined modulo a quantum of polarization, as discussed above. Specifically, Eq. (2) is modulo $e/A_{int}$,[26] where $A_{int}$ is the unit cell area of the interface. To resolve the ambiguity, we need to select a specific branch of the formal polarization for each material. We will do this by choosing a reference structure for each of the two materials, which allows connecting them at the specific interface under study.

For wurtzite, a layered hexagonal structure ($P6_3/mmc$) serves as a convenient reference structure.[27] It is derived from wurtzite by moving the cations into the plane of the anions, resulting in a centrosymmetric structure with zero $\mathbf{P}_f$.[27] The formal polarization of GaN is approximately linear as a function of the wurtzite internal parameter $u_{wz}$, which is defined as the ratio of the Ga–N bond length along the $c$ axis to the $c$ lattice parameter. The equilibrium structure of GaN has $u_{wz} = 0.377$, whereas layered hexagonal has $u_{wz} = 0.5$.

Similar to wurtzite [0001], rock salt in the [111] direction has alternating planes of cations and anions [see Fig. 1(c)]. However, the stacking sequence is "ABCABC," as opposed to "ABAB" in wurtzite. We follow a similar strategy to obtain a reference structure as we did for wurtzite: moving the cations into the plane of anions creates the face-centered cubic analog of the $P6_3/mmc$ GaN structure [Fig. 1(a)]. Unlike $P6_3/mmc$, this layered version of rock salt (space group $R3m$) is not centrosymmetric; however, since each layer is charge neutral, we expect the polarization of this structure to be very small, as will indeed be confirmed by explicit calculations.

To generalize $u_{wz}$ to a parameter that can be used with both wurtzite and rock-salt structures, we define $\delta$ as the ratio of the separation between planes of anions and cations to the separation between planes of cations; with this definition, $\delta = 1 - 2u_{wz}$ for the wurtzite structure. The layered structures, in which the anions and cations lie in the same plane, have $\delta = 0$. Ideal wurtzite has $\delta = 0.25$, and bulk

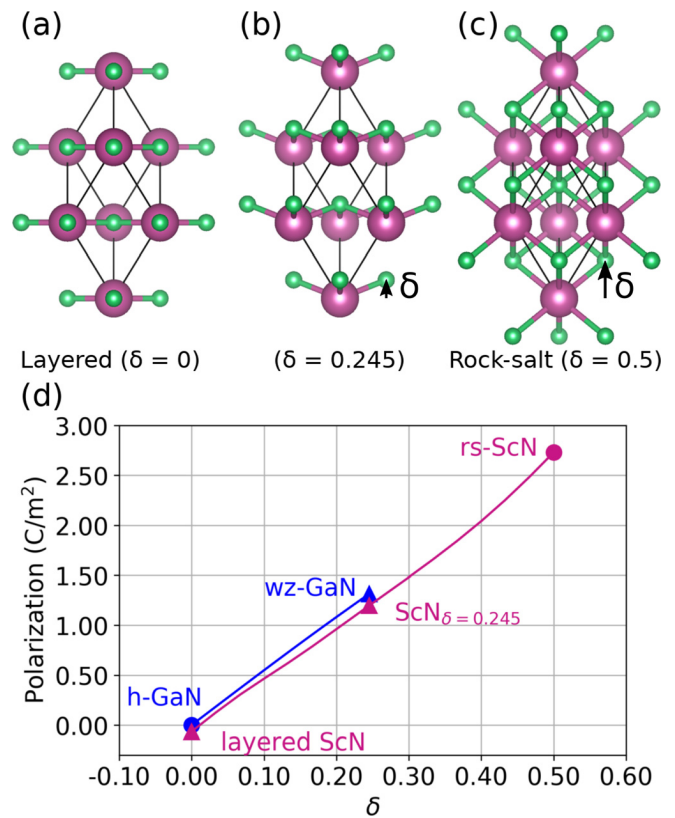

FIG. 1. ScN in the (a) layered, (b) $\delta = 0.245$, and (c) rock-salt crystal structures. Unit cells are indicated by black lines. (c) Calculated formal polarization as a function of the $\delta$ parameter. For ScN, the lattice parameters are held at the rs-ScN values, while for GaN, the lattice parameters are held at the wz-GaN values. The circles indicate centrosymmetric structures.

GaN has $\delta = 0.245$ along the $c$ axis. The rock-salt structure [Fig. 1(c)] has $\delta = 0.5$ along the [111] axis. We find that the formal polarization is approximately linear in $\delta$.

In Fig. 1(d), we plot a branch of the formal polarization of ScN and GaN as a function of $\delta$. The branches are chosen so that h-GaN has zero formal polarization and $\delta = 0$ ScN has a formal polarization close to zero. Our calculated formal polarization of rs-ScN along the $c$ axis is $\frac{ea_{rs}}{\Omega}||(\frac{1}{2},\frac{1}{2},\frac{1}{2})|| = 2.731$ Cm$^{-2}$, while the formal polarization of layered ScN is $-0.064$ Cm$^{-2}$. The calculated formal polarization of wz-GaN is 1.315 Cm$^{-2}$. According to the modern theory of polarization,[26] differences between formal polarization of two structures A and B are physically meaningful if a gap-preserving deformation exists between A and B; we have verified that the deformations as a function of $\delta$ are all gap-preserving, for both GaN and ScN.

As mentioned before, the lattice mismatch between rs-ScN and wz-GaN is quite small; however, if we consider ScN to be strained coherently to GaN, there will be a small piezoelectric contribution to the interface charge. Rock salt, being centrosymmetric, has no "proper" (in the sense of Ref. 28) piezoelectric response; however, strain in the (111) plane will dilute or concentrate the zero-strain formal polarization, and thus, there will be an "improper" contribution,[28] given by $-2\epsilon P_f^{ScN}$, where $\epsilon$ is the strain induced by the lattice mismatch in the (111) plane, $\epsilon = (a_{ScN}/\sqrt{2} - a_{GaN})/a_{GaN}$. We calculate this piezoelectric contribution to be $-0.058$ Cm$^{-2}$. The polarization charge at the interface can then be calculated using

$$\sigma_b = P_f^{GaN} - (1 - 2\epsilon)P_f^{ScN}, \quad (3)$$

and the predicted polarization difference of rs-ScN coherently strained to wz-GaN is $\sigma_b = 1.315 - (2.731 - 0.058) = -1.358$ Cm$^{-2}$, corresponding to a bound charge with a magnitude of $8.5 \times 10^{14}$ cm$^{-2}$. This value is over an order of magnitude larger than the bound charge at interfaces between, e.g., GaN and AlN.

In order to predict this bound charge based on differences between formal polarization values calculated for wz-GaN and rs-ScN, we applied the interface theorem [Eq. (2)]. The application of this theorem requires the interface to be insulating; we verify this by performing explicit superlattice calculations, in which the ScN layer is strained to the in-plane lattice parameters of GaN (but allowing all internal coordinates to relax). The layer-resolved density of states (DOS) is shown in Fig. 2(a). We immediately notice that strong electric fields are present. This is to be expected since large polarization charges are present at the interfaces. The fields are so large that the resulting voltage drop across the ScN layer exceeds the bandgap (no matter how thin we make this layer), leading to charge transfer between the interfaces. Though the presence of mobile charges at the interfaces results in a noninsulating cell, the essential feature in the layered DOS is the presence of a bandgap in each of the layers; therefore, the interface itself is insulating and our use of the interface theorem is justified.

In principle, the electric fields present in the superlattice should be quantitatively consistent with our calculation of bound interface charges based on differences in polarization quantities determined for bulk wz-GaN and rs-ScN. However, the presence of mobile charge complicates a direct comparison. It also precludes a direct verification of our choice of branches when calculating formal polarization differences based on bulk values [Fig. 1(d)]; indeed, the large polarization charge is on the same order as the quantum of interface polarization ($e/A_{int} = 1.803$ Cm$^{-2}$). We will address this by performing superlattice calculations between structures of GaN and ScN, which give rise to smaller polarization differences; these structures are based on different values of the interlayer spacing $\delta$ [Fig. 1(d)].

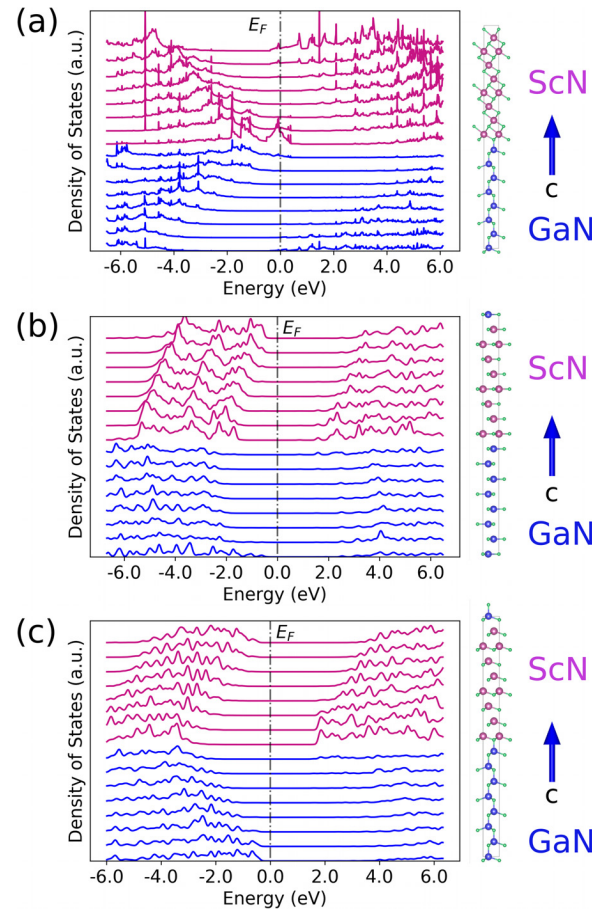

FIG. 2. Layer-resolved density of states for GaN/ScN superlattices with (a) wz-GaN/rs-ScN, (b) $\delta = 0$, and (c) $\delta = 0.245$. The superlattice structure is illustrated on the right, with Ga atoms in blue, Sc atoms in purple, and N atoms in green.

First, we construct a superlattice for the $\delta = 0$ structures, i.e., between h-GaN and the layered ScN structure [Fig. 1(a)]. Based on the formal polarization values in Fig. 1(d), and accounting for the improper piezoelectric effect (since ScN is strained to GaN), we expect a relatively small bound charge of 0.070 Cm$^{-2}$, at the interface. Indeed, in Fig. 2(b), we see that the electric fields are significantly smaller and breakdown is avoided, i.e., the entire superlattice is insulating. The fields in Fig. 2(b) are opposite in sign to those in Fig. 2(a) as the polarization of layered ScN is smaller than that of h-GaN, whereas the polarization of rs-ScN is larger than that of wz-GaN (see Fig. 1).

The electric fields extracted from the superlattice and the theory of linear dielectric media allow us to obtain the "zero field polarization difference" between the layers (i.e., removing the additional polarization from the dielectric response of the layers to the fields in the calculation, see Ref. 29). By Gauss's law, the electric displacement field is discontinuous across the interface, with the discontinuity being equal to the polarization difference,

$$\sigma_b = \varepsilon_0 \varepsilon_r^{\text{ScN}} \mathcal{E}_{\text{ScN}} - \varepsilon_0 \varepsilon_r^{\text{GaN}} \mathcal{E}_{\text{GaN}}, \quad (4)$$

where $\mathcal{E}$ is the electric field in each material, $\varepsilon_0$ is the permittivity of free space, and $\varepsilon_r$ is the relative permittivity. In the superlattice of the layered structures, the atomic positions are fixed; only the electrons screen the electric field, and the relevant permittivity is the clamped-ion dielectric constant. We calculate clamped-ion dielectric constants of 5.6 for h-GaN and 6.7 for layered ScN.

This procedure produces a bound polarization charge of 0.058 C m$^{-2}$, to be compared with the bound charge of 0.070 C m$^{-2}$ obtained from the difference in formal polarizations. Because these values are much smaller than the quantum of interface polarization, we can be confident that we have chosen the correct branch for the formal polarizations in Fig. 1(d).

As an additional test, we repeat the procedure for an interface between wz-GaN and ScN with $\delta = 0.245$, with respective clamped-ion dielectric constants of 5.1 and 6.5. The ScN layer is strained to the in-plane lattice parameters of GaN. The results are shown in Fig. 2(c). An analysis of the electric fields allows us to extract a polarization charge of 0.121 C m$^{-2}$ at the interface. This is to be compared with a value of 0.126 C m$^{-2}$ derived from the formal polarization difference. This again confirms that we have made the correct choice of branch for the polarization charge at the (0001)wz-GaN/(111)rs-ScN interface.

The procedure for determining polarization charge at the (0001)wz-GaN/(111)rs-ScN interface can equally be applied to similar interfaces between rs-ScN and other nitride semiconductors (wz-AlN or wz-InN) for which polarization values are known.[27] Using the value of 2.731 C m$^{-2}$ for the formal polarization of rs-ScN and Eq. (3) to include the effect of the piezoelectric field, polarization charges between ScN and any wurtzite nitride can then be computed.

In the absence of free carriers, breakdown will occur if the potential drop over the ScN layer is larger than the bandgap of ScN. For a rs-ScN layer grown on semi-infinite wz-GaN, we can apply Eq. (4), setting $\mathcal{E}_{\text{GaN}} = 0$. The electric field in the ScN layer will be $|\sigma_b|/\varepsilon_0 \varepsilon_r^{\text{ScN}} = 6.1$ GV/m (using $\sigma_b = -1.358$ C m$^{-2}$ and our calculated static dielectric constant of 25.0). The critical layer thickness for breakdown to occur is then given by $d_c = E_g/e\mathcal{E} = 1.5$ Å, approximately one monolayer. Therefore, for any thickness of the ScN layer, a 2DEG forms when ScN is grown at the (000$\bar{1}$) GaN interface and a 2DHG forms at the (0001) GaN interface.

For finite-thickness layers, as the thickness of the GaN and ScN layers increases, increasing numbers of holes or electrons will appear at the interfaces to compensate the bound polarization charge. As the thicknesses of the GaN and ScN layers increase to infinity, the fields in the layers will vanish, and the free-carrier density at the interface will equal the bound polarization charge density ($8.5 \times 10^{14}$ cm$^{-2}$).

The hole and electron gases at the GaN/ScN interfaces may be useful for contacts or as current spreading layers, due to the extremely high carrier concentrations. p-GaN/ScN/n-GaN [grown in the (0001) direction] tunnel junctions are another attractive application. Current tunnel junctions make use of the smaller bandgap InGaN interlayers to reduce the effective barrier and increase the polarization field across the p-n junction to reduce the junction width.[11,12] ScN is a promising alternative, with a higher polarization field, smaller bandgap, and much smaller strain.

In summary, we have demonstrated a large polarization difference between ScN and GaN. The polarization difference between rs-ScN and wz-GaN is $-1.358$ C/m$^2$ when the formal polarizations are appropriately referenced. This polarization difference produces extremely large bound charges and electric fields, which can be exploited for high-density electron and hole gases with concentrations up to $8.5 \times 10^{14}$ cm$^{-2}$, in tunnel junctions.

This work was supported by the Air Force Office of Scientific Research under Award No. FA9550-18-1-0237. Computational resources were provided by the Department of Defense High Performance Computing Modernization Program, by the Extreme Science and Engineering Discovery Environment (XSEDE), which was supported by National Science Foundation Grant No. ACI-1548562, and by the Center for Scientific Computing, which was supported by the California NanoSystems Institute and the Materials Research Science and Engineering Center (MRSEC) at UC Santa Barbara through Nos. NSF DMR-1720256 and NSF CNS-1725797.